\documentclass{article}% Chicago-based Humanities Reference Style

\usepackage[inline]{enumitem}
\usepackage{graphicx}%
\usepackage{multirow}%
\usepackage{amsmath,amssymb,amsfonts}%
\usepackage{amsthm}%
\usepackage{mathrsfs}%
\usepackage[title]{appendix}%
\usepackage{xcolor}%
\usepackage{textcomp}%
\usepackage{manyfoot}%
\usepackage{booktabs}%
\usepackage{algorithm}%
\usepackage{algorithmicx}%
\usepackage{algpseudocode}%
\usepackage{listings}%
\usepackage{booktabs}
\usepackage{subcaption}
\usepackage{array}

\usepackage{arxiv}
\usepackage[utf8]{inputenc} % allow utf-8 input
\usepackage[T1]{fontenc}    % use 8-bit T1 fonts
\usepackage{hyperref}       % hyperlinks
\usepackage{url}            % simple URL typesetting
\usepackage{booktabs}       % professional-quality tables
\usepackage{amsfonts}       % blackboard math symbols
\usepackage{nicefrac}       % compact symbols for 1/2, etc.
\usepackage{microtype}      % microtypography
\usepackage{natbib}
\usepackage{doi}

\title{An Open-Source, Autonomous Platform for High-Resolution Energy Monitoring in Manufacturing}

\author{ \href{https://orcid.org/0000-0002-4890-7264}{\includegraphics[scale=0.06]{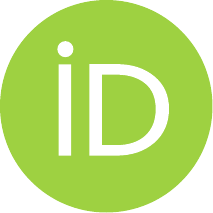}\hspace{1mm}Vignesh Selvaraj}\\
	Department of Mechanical Engineering\\
	University of Wisconsin-Madison\
	Madison, WI 53706 \\
	\texttt{vselvaraj@wisc.edu} \\
	\And
	\href{https://orcid.org/0000-0002-9542-0803}{\includegraphics[scale=0.06]{orcid.pdf}\hspace{1mm}Aditya Nagaraj} \\
	Department of Mechanical Engineering\\
	University of Wisconsin-Madison\
	Madison, WI 53706 \\
	\texttt{adityanagaraj@outlook.com} \\
	\And
    \href{https://orcid.org/0009-0008-4203-1995}{\includegraphics[scale=0.06]{orcid.pdf}\hspace{1mm}Shengyuan Zhang} \\
	Department of Mechanical Engineering\\
	University of Wisconsin-Madison\
	Madison, WI 53706 \\
	\texttt{szhang886@wisc.edu} \\
    \And
    \href{https://orcid.org/0009-0000-3715-5143}{\includegraphics[scale=0.06]{orcid.pdf}\hspace{1mm}Sina Sadeghian} \\
	Department of Mechanical Engineering\\
	University of Wisconsin-Madison\
	Madison, WI 53706 \\
	\texttt{sina.sdgn@gmail.com} \\
    \And
    \href{https://orcid.org/0000-0002-3550-3711}{\includegraphics[scale=0.06]{orcid.pdf}\hspace{1mm}Sangkee Min} \\
	Department of Mechanical Engineering\\
	University of Wisconsin-Madison\
	Madison, WI 53706 \\
	\texttt{sangkee.min@wisc.edu} \\
}

\renewcommand{\shorttitle}{Autonomous Energy Monitoring System}

\begin{document}

\maketitle

%%==================================%%
%% Abstract %%
% Need to be fixed - Currently high-level
%%==================================%%
\begin{abstract}
    High-resolution energy data is increasingly central to Industry 4.0, where electrical signals such as three-phase voltage and current carry rich information about machine condition, tool wear, and process dynamics. Capturing this information in practice remains difficult: commercial power analysis are largely proprietary, offer limited or no access to high-sampling rate data for transient analysis, restrict access to raw waveform data, and offer no customization, while general-purpose open hardware lacks the front-end accuracy, isolation, and robustness required for industrial measurement. This paper presents Autonomous Energy Monitoring System (AEMS), an open-source, low-cost, and modular platform supported by a host, edge-gateway, and optional cloud software stack that enables autonomous, long-duration acquisition independent of a continuously connected host and thereby closes this gap by combining research-grade fidelity with industrial deployability. The system acquires three-phase voltage and current through an isolated front-end and a 24-bit, simultaneously sampling analog-to-digital converter, managed by a dual-core architecture that separates deterministic acquisition and on-board logging from host communication and control. Industrial interfaces (Ethernet, RS-485/Modbus, and BLE) together with hardware-level synchronization enable scalable, time-aligned acquisition across multiple machines, supported by a complete host, edge-gateway, and optional cloud software stack. We validate the platform on a three-axis CNC machining center, where it resolves spindle, feed-drive, rapid-traverse, and material-removal energy states and detects feed-rate changes as small as 50 mm/min. By releasing the full hardware and firmware openly, this work aims to democratize access to high-fidelity energy monitoring for both researchers and small and medium-sized manufacturers.
\end{abstract}

\keywords{Energy Monitoring, Open-source hardware, Condition Monitoring, Industry 4.0, Data acquisition}



\maketitle

\section{Introduction}\label{intro}
% 1. Motivation: energy monitoring requirement in manufacturing
% 2. Challenges in existing solutions (cost, scalability, accessibility)
% 2.1 The need for open source in smart sensor systems
% 3. Contributions of this work (open-source, modular h/s, case study)
% 4. Paper organization

The transition towards Industry 4.0 has created an urgent need for intelligent, connected, and data-driven manufacturing systems. An integral part of this transformation is the ability to capture high-fidelity data from machines and processes. The data enables monitoring, optimization, and predictive decision-making \citep{gao2020big}. For instance, in manufacturing, energy consumption signals such as current and voltage carry rich information about machine condition, tool wear, process dynamics, and efficiency. When measured at sufficiently high sampling rates, these signals reveal transient behaviors that are critical for understanding and controlling complex production systems.

However, existing commercial Data Acquisition (DAQ) and energy monitoring solutions are often proprietary, expensive, and limited in configurability. Most are designed for aggregated power monitoring or billing applications and therefore operate at low sampling frequencies, offer restricted access to raw data, and rely on closed communication protocols. Such constraints hinder the development of machine-learning and AI models, which require large amounts of synchronized, high-resolution data for training and validation. Furthermore, the lack of modularity and scalability in conventional DAQ systems limits their adaptability across different machines, processes, sensor modalities, and deployment scales, both from laboratory research to full-scale industrial environments.

The development of modular, scalable, cost-effective, and open-source DAQ systems is therefore essential for realizing the full potential of Industry 4.0. Open-hardware platforms provide transparency, interoperability, and cost efficiency while empowering researchers and manufacturers to customize sensing, synchronization, and communication capabilities to their specific needs. Scalable architectures enable distributed, multi-machine data collection, while high-speed, high-precision sampling supports the creation of digital twins and AI-based analytics for condition monitoring, and energy optimization. By democratizing access to advanced measurement infrastructure, such systems help bridge the technological divide between large corporations and small and medium-sized enterprises (SMEs).

In this context, this paper presents Autonomous Energy Monitoring System (AEMS) \footnote{Open-source hardware, firmware, and software:
\url{https://github.com/vigneshuw/aems} (archived at
\href{https://doi.org/10.5281/zenodo.21383443}{10.5281/zenodo.21383443}).}, an open-source, low-cost, and modular hardware–software platform for high-resolution energy data acquisition in manufacturing. The autonomous here refers to the system's ability to enable long-term, unattended data acquisition that continues through host or network outages suitable deployment in real manufacturing industries. The contributions of this work are threefold. First, a custom hardware design that acquires three-phase voltage and current through an isolated front-end and a 24-bit, simultaneously sampling  Analog-to-Digital Converter (ADC) at upto 32 kS/s per channel, with on-board logging and industrial communication interfaces, and supports hardware-level synchronization for time-aligned, multi-machine acquisition. Second, a complete open software stack, a host communication library, an edge-gateway daemon, and an optional cloud interface, spanning deployment from direct laboratory experimentation to autonomous, plant-scale monitoring. Third, an experimental case study on a three-axis CNC machining center demonstrating that the system resolves spindle, feed-drive, rapid-traverse, and material-removal energy states and detects feed-rate changes as small as 50~mm/min. The full hardware and firmware are released openly to support reproducibility and extension.

The remainder of the paper is organized as follows. Section \ref{sec:bg} provides a literature review on the existing technologies and DAQ systems for energy consumption monitoring for manufacturing industries. Section \ref{sec:aems} describes the design and implementation of Autonomous Energy Monitoring System, including its hardware, firmware, and software architecture. Section \ref{sec:case_study} presents a case study illustrating system performance in controlled experiments. Finally, Section \ref{sec:discussion} discusses the key findings, and Section~\ref{sec:conclusion} concludes the paper and outlines directions for future work.

\section{Related Work}\label{sec:bg}
% 1. Energy monitoring approaches in manufacturing
% 2. Comparison with other available hardware - industrial vs open-source
% 3. Gaps addressed in this work

\subsection{Energy Monitoring in Manufacturing}

Monitoring energy consumption in manufacturing industries has been extensively studied across multiple scales, ranging from facility-level assessments to detailed process- and machine-level analyses aimed at improving overall energy efficiency \citep{denkena2020energy}. Manufacturing alone accounts for 84\% of the industrial sector’s energy-related CO\textsubscript{2} emissions and 90\% of its total energy consumption \citep{duflou2012towards}, underscoring the need for effective monitoring strategies. Prior work has emphasized that meaningful reductions in energy usage require visibility across the full hierarchy of manufacturing operations. For example, \cite{rahimifard2010minimising} highlight the importance of monitoring energy from the enterprise level down to the machining state level, arguing that significant improvements are achievable only through a “Design for Energy Minimization’’ philosophy that depends on high-resolution information across multiple organizational layers. Manual approaches for energy consumption monitoring at each level are impractical, hence, automation is a vital step in characterizing the energy consumption of complex systems as described by \cite{vijayaraghavan2010automated}. Within this work the authors describe an Event Streaming Processing (ESP) technique to automate the monitoring and analysis of energy consumption. The ESP technique necessitates monitoring of energy consumption at different scales, from Tool-Chip interface to manufacturing supply chain, hence, the need for cost-effective, open-source, and scalable energy meters is evident.

Commercially available power meters have been widely used for industrial energy monitoring \citep{o2013industrial}. However, these systems often fall short of the needs of modern manufacturing and research communities. They tend to be costly, offer limited configurability, and restrict access to raw high-frequency data—an essential requirement for transient analysis, machine-level diagnostics, and AI/ML applications. Selecting a suitable power meter is further complicated by practical considerations such as sampling rate, accuracy, resolution, and compatibility with industrial communication protocols. Seamless integration with existing supervisory and data management systems frequently becomes a limiting factor, particularly for facilities pursuing Industry 4.0 initiatives.

\cite{o2013industrial} categorize industrial energy monitoring into three hierarchical levels: first-order (facility level), second-order (process level), and third-order (machine level). This type of categorization is common and was also found in other works \citep{kara2011electricity}. Sampling rate serves as a key differentiator among these levels. Facility-level monitoring may only require a few samples per hour to track aggregate consumption, whereas machine-level monitoring demands significantly higher sampling frequencies to capture short-duration transients and dynamic load variations inherent to manufacturing processes. Although the authors describe several high-end commercial meters capable of high sampling rates, such systems typically do not provide continuous access to raw waveform data, are not cost-effective for large-scale deployment, and rely on fixed, proprietary architectures. These limitations severely constrain their use in advanced manufacturing analytics, including the development of AI/ML models, high-frequency transient diagnostics, and large-volume data collection for research. 

\subsection{High-Fidelity Sensing: Applications and Constraints}

The value of high-resolution, machine-level energy data is most evident in process and tool condition monitoring, where electrical signals provide an indirect but non-intrusive measure of the cutting process. As surveyed by \cite{teti2010advanced}, spindle motor current and power are well-established modalities for monitoring tool wear, breakage, and process anomalies, favored in industrial settings precisely because they require no additional in-process sensors. Building on this principle, \cite{drouillet2016tool} showed that the root-mean-square of spindle power is sensitive to tool wear and can be combined with a neural network to predict remaining useful life in milling, while \cite{corne2017study} used spindle power with neural networks for real-time tool-wear and breakage prediction during drilling. A common thread across these studies, however, is a reliance on the machine's internal drive signals or on external commercial instruments: the former are convenient but aggregated and bandwidth-limited, exposing processed quantities rather than raw waveforms and in some cases may not be readily accessible, while the latter reintroduce the cost and closed-data constraints discussed above. The diagnostic value of electrical signatures is thus well established, but realizing it at the fidelity that data-driven methods demand depends on measurement hardware that provides synchronized, high-rate, raw access to the underlying voltage and current, the capability conventional instruments withhold.

A complementary direction is the disaggregation of energy consumption through non-intrusive load monitoring (NILM), introduced by \cite{hart1992nonintrusive}, which resolves aggregate electrical measurements into the contributions of individual loads from a single measurement point. The fidelity of disaggregation improves with richer electrical features: low-frequency RMS quantities suffice for coarse load recognition, whereas high-frequency information such as harmonic content and switching transients substantially improves the separation of similar loads \citep{laughman2003power, zeifman2011nonintrusive}. Although NILM is well developed for residential and building applications, its extension to the industrial domain, where loads are fewer but operationally far more complex, remains comparatively early-stage \citep{yaniv2025advances}. In this setting, high-resolution, synchronized, per-machine measurements of the kind targeted in this work serve a dual role: as a direct monitoring capability in their own right, and as the high-fidelity, labeled ground-truth data required to train and validate disaggregation and other data-driven energy models.

In recent years, emerging technologies such as the Internet of Things (IoT) have enhanced process monitoring in manufacturing industries. An area in particular that can reap significant benefits is energy consumption monitoring \citep{shrouf2015energy}. The work provides framework for integrating IoT with company's existing IT systems. Beyond hardware constraints, several studies have underscored the need for scalable and practical sensing deployments in real industrial environments. \cite{lu2009online} proposed a wireless sensor network (WSN) for remote energy monitoring and fault diagnostics, demonstrating the value of electrical signatures in equipment condition monitoring while emphasizing the challenges of DAQ and WSN deployment in factories. Similarly, \cite{o2012implementation} stress the importance of sophisticated energy metering systems in reducing operational costs and meeting environmental targets, noting that widespread adoption is hindered by the trade-off between monitoring capability and affordability. While capturing transient energy events is highly desirable, the escalating cost of meters capable of doing so makes facility-wide deployment economically impractical.

These observations reflect a recurring theme in the literature: industrial energy monitoring sits at the intersection of hardware expense, measurement fidelity, and communication constraints. Studies consistently point to a tension between the need for high-resolution sampling, to detect short-lived transients and dynamic load variations, and the prohibitive cost or rigidity of advanced metering systems. At the same time, although wireless sensor networks provide valuable flexibility, their limited data throughput forces difficult design compromises between centralized analysis of raw data and localized processing of only selected features or events.

\subsection{Open-Source Hardware and the Research Gap}

In parallel with these challenges, recent work has explored the role of Open-Source Hardware (OSH) in manufacturing. \cite{stange2023open} report that while OSH is currently most prevalent in the hobbyist community with the impact in manufacturing domain being as low as 4.6\%. OSH has also demonstrated substantial benefits in reducing laboratory equipment costs \citep{pearce2014cut}. \cite{pearce2020economic} demonstrates that open-source approaches can lead to 87\% baseline savings compared to proprietary scientific tools. More importantly, OSH offers a powerful mechanism for accelerating innovation in manufacturing by improving the transfer of knowledge from universities to industry. Collaborative OSH projects enable shorter development cycles and allow research prototypes to align more rapidly with industrial needs. Such approaches are particularly valuable for small and medium-sized enterprises (SMEs), which often lack the resources for extensive research investment but can enhance their innovation potential through open collaboration \citep{sedita2016role}.

Open-source in the domain of Industry 4.0 can be hardware or software-based. With the advent of hardware platforms such as Arduino, Raspberry Pi, etc., several evaluation boards were developed and used extensively in IoT applications \citep{krushinitskiy2013review}. However, the industrial use of such hardware is still disputed, although several companies have developed many industrial versions of the Raspberry Pis. \cite{martikkala2021trends} identified several challenges that can hinder the adoption of open-source solutions for the Industrial Internet of Things (IIoT), the key elements being interoperability, standardization, reliability, regulatory compliance and support, including the fact that there is an urgent need for open-source solutions in manufacturing. 

Within energy monitoring specifically,  a small number of open-source meters have begun to address the cost and accessibility limitations of proprietary instruments. \cite{oberloier2018open}  presented a fully open-source, low-cost power monitoring system built around modular logging nodes, and \cite{pocero2017open} developed open-source IoT metering devices for energy-efficient buildings. More recently, openZmeter \cite{viciana2018openzmeter} demonstrated an open-hardware platform for energy and power-quality measurement in low-voltage systems. These efforts confirm both the feasibility and the value of open energy instrumentation. They are, however, oriented toward facility-, building-, and grid-level applications and consequently operate at sampling rates and with architectures suited to those tasks. They are not designed for the high-rate, simultaneously sampled, multi-channel acquisition required to resolve the short-duration transients and per-machine energy states that characterize manufacturing processes, nor for the synchronized, multi-machine deployment needed at the plant level. The open instrumentation that exists, in other words, addresses a different measurement regime than the one machine-level manufacturing analytics demands.

Despite the practical and technical challenges, the overall trend in the field points toward increasingly intelligent, scalable, and non-intrusive monitoring systems. Emerging technologies such as the Industrial Internet of Things (IIoT), Non-Intrusive Load Monitoring (NILM), and advanced WSNs are shaping a new generation of sensing architectures. Together, these developments highlight a growing need for affordable, modular, high-fidelity, and open-source data acquisition systems capable of delivering raw, high-frequency energy data suitable for industrial deployment and advanced research applications such as AI/ML, digital twins, and predictive maintenance. This unmet need provides the central motivation for the development of Autonomous Energy Monitoring System (AEMS).

% Summary Paragraph
% 1. Importance of energy monitoring in industries on a large scale
% 2. Hardware constraints, high sampling rate, costs etc., Limied hardware specs
% 3. Communication protocols and impact of Data rate bits o2012implementation
% 4. The need for communication between the boards.

\section{Autonomous Energy Monitoring System (AEMS)}\label{sec:aems}

\subsection{Design Objectives and Requirements}
The AEMS system was designed considering the following requirements:
\begin{enumerate*}[label=(\roman*)]
    \item high sampling rate,
    \item measurement accuracy,
    \item modularity, 
    \item synchronous data acquisition across machines,
    \item compatibility with industry protocols, 
    \item edge processing,
    \item low cost and
    \item autonomous long duration acquisition capable of handling temporary host connection loss.
\end{enumerate*}
The high sampling rate and access to the raw data for downstream processing enable capturing the energy transients that occur during equipment state changes such as spindle start-up, change in RPM, engagement with the workpiece, etc. With the ability to measure accurately, subtle variations in power consumption can be reliably detected. This can provide early indications of tool wear, inefficiency, or malfunction. The modular hardware design allows the system to scale and interface with other edge devices in the factory, supporting broader monitoring networks and system upgrades. The ability to synchronously collect energy consumption data across multiple machines enables the correlation of events at the plant level, giving insights into machine interactions and production bottlenecks. Compatibility with widely adopted industrial communication protocols ensures seamless integration into existing factory systems. Capable hardware enabling data processing at the edge to reduce network throughput and support local decision-making. Finally, the emphasis on low-cost and open source accessibility reduces the barrier to adoption, making advanced energy monitoring feasible not only for large manufacturers but also for small and medium-sized enterprises and research labs.

\subsection{Hardware Design}
The hardware design was guided by the requirements of accuracy, modularity, and robustness for long-term deployment in industries. At the high level, the system consists of three main components: a sensing and measurement front-end, a data acquisition core with the capability of processing and logging, and a communication subsystem with wired and wireless capabilities. Together, these elements form a modular architecture that can be scaled to monitor single machines or extended across multiple devices in a factory. 

\begin{table}[htbp]
\centering
\caption{Summary of the proposed energy monitoring hardware specifications.}
\label{tab:ems_hardware_specs}
\begin{tabular}{p{0.30\linewidth} p{0.60\linewidth}}
\toprule
\textbf{Parameter} & \textbf{Specification} \\
\midrule
Voltage measurement & Three-phase line-to-neutral input, up to \(230~\mathrm{V_{RMS}}\) per channel (nominal \(400~\mathrm{V_{LL}}\) system) \\
Current measurement & Three-phase current input using \(333~\mathrm{mV}_{\mathrm{RMS}}\) current transformers \\
Sampling capability & Simultaneous multi-channel sampling at a max of \(32~\mathrm{kS/s}\) per channel \\
Local data storage & \(4~\mathrm{GB}\) on-board eMMC storage with microSD card interface \\
Wired communication & \(100\mathrm{BASE\text{-}TX}\) Ethernet and RS485 interface \\
Wireless communication & Bluetooth Low Energy (BLE) interface \\
\bottomrule
\end{tabular}
\end{table}

The sensing front-end was designed to capture both voltage and current. The voltage measurement was achieved using a precision voltage divider and the current measurement was achieved using precision opamp (operational amplifier). The detailed specifications of the hardware can be seen in Table \ref{tab:ems_hardware_specs}. Additional surge protection was incorporated to protect circuits against industrial power disturbances and high voltage. These choices allow for reliable capture of both steady-state power draw and transient events associated with machine start-up, load changes, or tool engagement.

The data acquisition core is built around the ADS131M08 from Texas Instruments, a 24-bit, 32-kSPS, 8-channel simultaneous-sampling analog-to-digital converter (ADC). The use of simultaneous sampling is essential for accurate instantaneous power measurements, as it allows current and voltage channels to be captured without timing skew. The relatively high sampling rate further enables the system to resolve short-lived load transients and rapid state changes in manufacturing equipment. Data acquisition and system control are managed by a dual-core STM32H745ZITx microcontroller from STMicroelectronics, which oversees buffering, local storage, and communication with higher-level devices. This MCU was selected for its dual core computing, real-time processing capability, support for multiple industrial communication protocols, and sufficient bandwidth to handle continuous raw data streaming to edge devices on the factory floor.

The proposed hardware provides multiple interfacing options to support both wired and wireless communication. A Fast Ethernet interface enables TCP/IP or UDP-based communication with the board, supporting configuration, command exchange, and high-throughput data streaming. An RS-485 interface is included for industrial communication and can be configured for Modbus-RTU operation, allowing integration with PLCs and other industrial control systems. In addition, Bluetooth Low Energy (BLE) provides a wireless interface for board-level command, control, and monitoring. The hardware also integrates on-board eMMC storage and a microSD card interface for edge data logging. This local storage can serve multiple purposes, including raw data acquisition, temporary buffering, long-duration machine/process monitoring, and storage of machine learning model parameters for edge inference. Furthermore, on-board storage enables high-sampling-rate data acquisition in cases where real-time network streaming is limited by communication bandwidth. An expansion connector is also provided to support communication with additional energy monitoring modules. This connector can be used for hardware-level synchronization through general-purpose input/output (GPIO) signals, enabling coordinated multi-board acquisition for distributed or multi-point energy monitoring applications.

Power management was a critical consideration in the design. Onboard regulation generates separate isolated rails for the analog and digital domains, thereby minimizing coupling noise and improving measurement fidelity. A high-precision low-dropout regulator (LDO) provides a stable reference voltage for the ADC, ensuring consistent accuracy. In addition, distinct voltage domains were allocated to different subsystems to further suppress interference and enhance overall system robustness.

\begin{figure}[htbp]
    \centering
    \includegraphics[width=\textwidth]{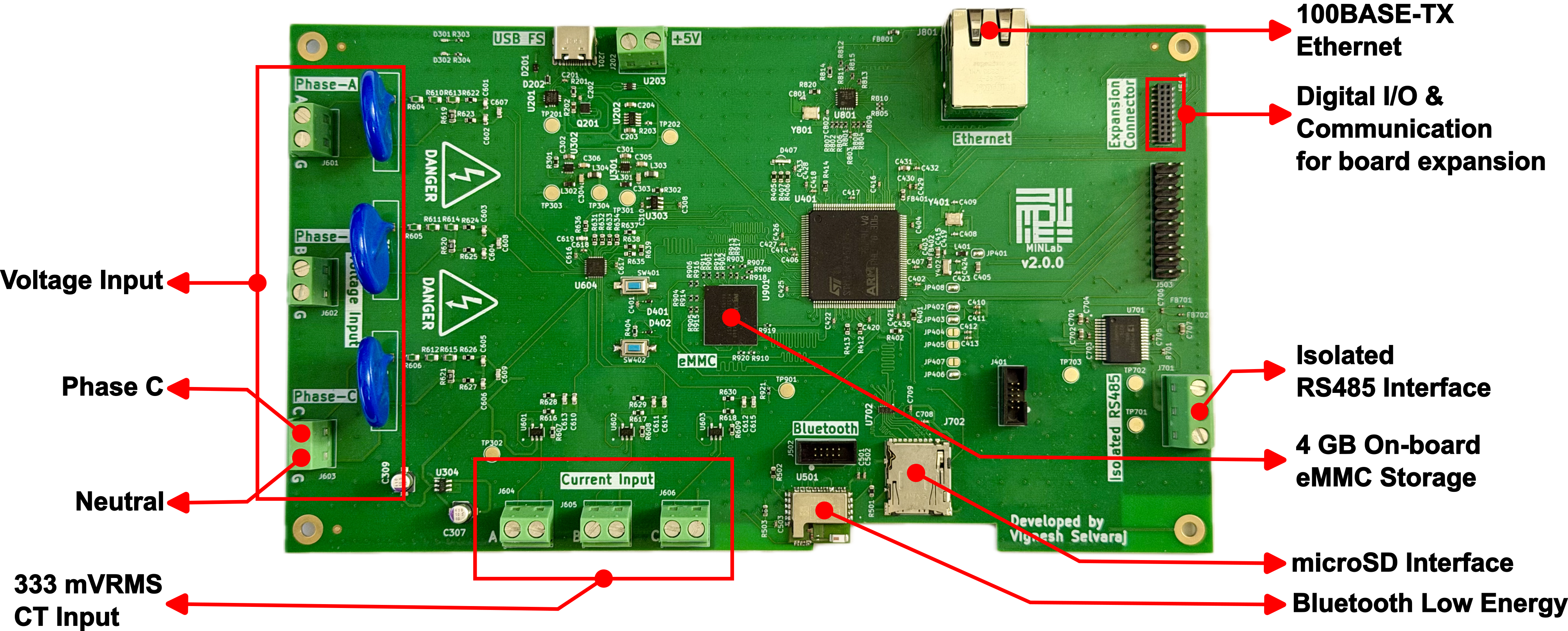}
    \caption{Overview of the energy monitoring hardware.}
    \label{fig:ems_hardware}
\end{figure}

Finally, the physical and PCB design of the system reflects its intended industrial application. Impedance matching was performed to ensure the signal integrity at high sampling rates for DAQ and logging to SD card. The modular form factor also supports future expansion, allowing additional sensing channels or communication interfaces to be added without redesigning the core system. A snapshot of the designed hardware along with its various components highlighted can be seen in Figure \ref{fig:ems_hardware}.

Through these design choices, the hardware achieves a balance of precision, robustness, and flexibility that sets it apart from conventional energy meters. The system was deliberately engineered to serve dual purposes: providing researchers with high-fidelity measurements for controlled laboratory studies while also meeting the reliability and durability requirements necessary for long-term deployment in industrial environments. Figure \ref{fig:ems_hardware} labels the different components of the hardware that showcase its capabilities.

\subsection{Software Design}
\label{subsec:software_design}

The software stack developed for the proposed energy monitoring hardware consists of two major components: embedded firmware running on the energy monitoring board and host-side software used to configure, control, monitor, and retrieve data from the device. Both components were designed as open-source software to support reproducibility, customization, and integration into laboratory and industrial monitoring workflows. The embedded firmware implements deterministic data acquisition, local storage, inter-core coordination, diagnostics, and board-level communication. The host-side software provides a higher-level interface for command execution, data streaming, file retrieval, metadata generation, and autonomous operation.

\subsubsection{Embedded Firmware Architecture}
\label{sec:embed_firmware}

The firmware was implemented on the dual-core STM32H745 microcontroller using a separation of responsibilities between the two processing cores. The Cortex-M4 core acts as the deterministic data-plane processor and is responsible for analog front-end acquisition, ADC servicing, sample buffering, and local storage operations. It interfaces with the ADS131M08 analog-to-digital converter, manages the SDMMC/eMMC storage interface, and executes the data acquisition state machine. The Cortex-M7 core acts as the supervisory and communication processor. It runs FreeRTOS and LwIP, maintains the host communication link, receives and dispatches host commands, and forwards acquisition and file-system requests to the Cortex-M4 core through OpenAMP/RPMsg.

\begin{figure}[htbp]
    \centering
    \includegraphics[width=0.8\textwidth]{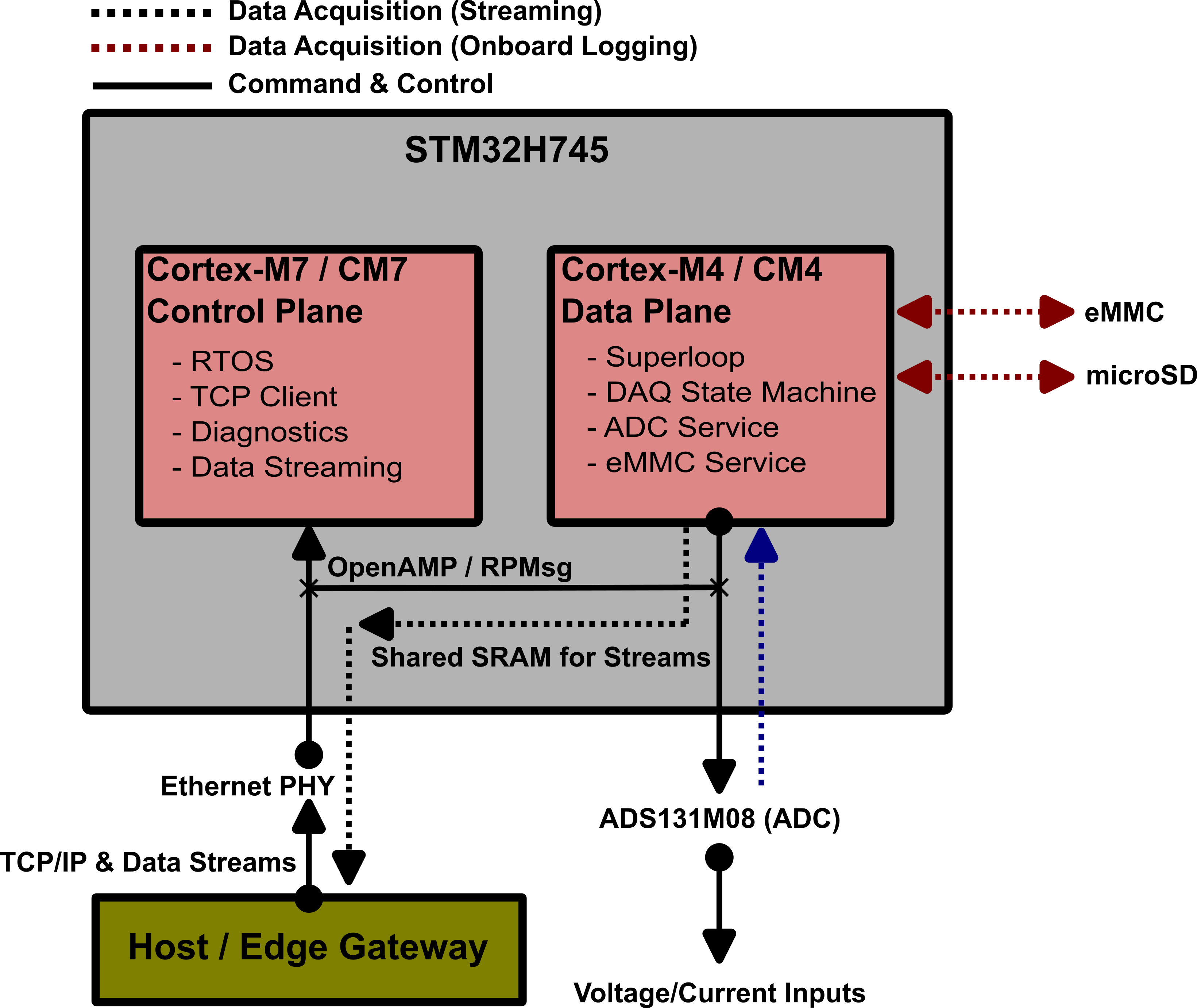}
    \caption{Embedded firmware architecture of the proposed energy monitoring hardware. The Cortex-M4 core performs deterministic acquisition, transfer, and storage, while the Cortex-M7 core manages host communication, command dispatch, diagnostics, and data streaming.}
    \label{fig:aems_embedded_firmware_architecture}
\end{figure}

This dual-core organization isolates timing-sensitive acquisition and storage tasks from host communication and command-processing tasks. The Cortex-M4 core owns the eMMC and file-system operations, while the Cortex-M7 core accesses stored data only through Cortex-M4-mediated OpenAMP services. This ownership model reduces file-system concurrency issues and ensures that storage timing remains under the control of the acquisition core. The embedded firmware architecture can be seen in Figure \ref{fig:aems_embedded_firmware_architecture}.

For the basic data logging workflow, the board operates as a remotely configurable data acquisition device. The host-side application defines acquisition parameters such as sampling configuration, channel selection, acquisition mode, and output file name. These parameters are transmitted to the board through the host communication interface, parsed by the Cortex-M7 control task, and forwarded to the Cortex-M4 acquisition service through the inter-core command path. The Cortex-M4 firmware validates the request, prepares the ADC and storage subsystem, and transitions the acquisition state machine from the idle state to the acquisition state.

During acquisition, ADC data-ready events trigger transfers from the ADS131M08. The received samples are converted into internal sample frames, aggregated into blocks, and written to local storage when the firmware is operating in logging mode. The acquisition firmware follows a cooperative state-machine design with idle, preparing, acquiring, stopping, finalizing, and error states. Interrupt routines are kept lightweight and are used primarily to signal ADC data-ready events, while sample packaging, buffering, and file writing are performed in the main firmware execution context.

The local storage pathway enables high-rate data collection at the edge. In the implemented eMMC logging mode, selected ADC channels are packed as signed 32-bit little-endian values and written directly to binary log files. This allows the board to continue acquiring data even when continuous real-time streaming is limited by network bandwidth or host availability. The same storage subsystem supports file counting, file listing, file-size queries, file deletion, and file streaming back to the host.

The data streaming pathway enables high-rate data collection and transfer over Fast Ethernet. In the implemented streaming pathway selected ADC channels are packed as signed 32-bit little-endian values and streamed over TCP/IP. This enables the board to stream the data to a listening device without having to locally store data onboard. Additionally, multiple boards can be connected together using a simple unmanaged switch enabling data stream from multiple boards, enabling synchronous multi-device DAQ across the factory floor. 

\subsection{Host Interface}
\label{subsec:host_interface}

A host-side software stack was developed to provide a high-level interface for configuring, controlling, monitoring, and retrieving data from the proposed energy monitoring hardware. The host interface was designed to support multiple deployment modes, ranging from direct laboratory use with a desktop computer to autonomous edge deployment using a Raspberry Pi and optional cloud connectivity. At a high level, the host interface consists of three components: a host communication library, a Raspberry Pi edge gateway, and an optional cloud interface.

\subsubsection{Host Communication Library}
\label{sec:host_comm}

\begin{figure}[htbp]
    \centering
    \includegraphics[width=0.9\textwidth]{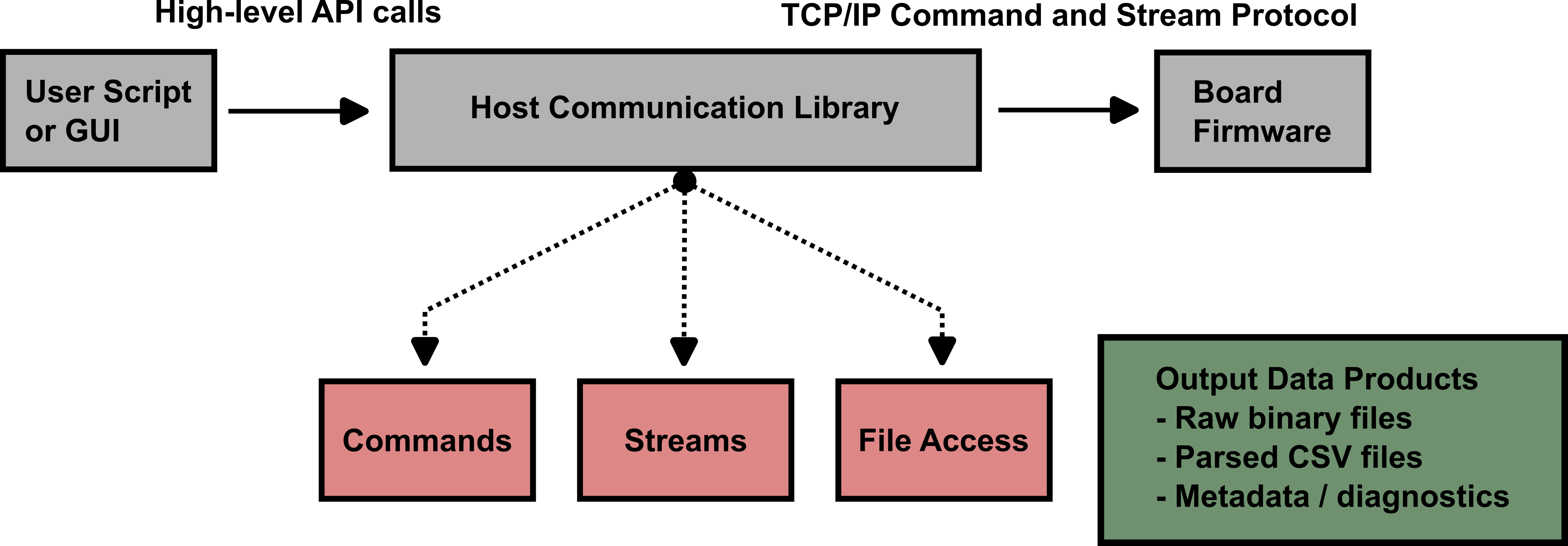}
    \caption{High-level workflow of the host communication library used for board configuration, command execution, data streaming, file retrieval, and offline data conversion.}
    \label{fig:aems_host_communication_library}
\end{figure}

A host communication library was developed to provide a reusable abstraction over the AEMS board communication protocol. The library enables users to interact with one or more boards through high-level software commands without directly implementing the low-level command, response, and streaming protocol. Boards are identified by their IP addresses, allowing both single-board and multi-board workflows. The workflow can be seen in Figure \ref{fig:aems_host_communication_library}.

The library provides high-level functions for board connection management, command execution, data streaming, response parsing, and data conversion. Supported operations include heartbeat checks, diagnostic queries, DAQ status requests, eMMC logging control, file listing, file deletion, file-size queries, file retrieval, and live DAQ streaming. The response-parsing layer converts board acknowledgments and diagnostic responses into structured software objects, making the interface suitable for integration with custom monitoring, automation, and data-processing applications.

The library also includes utilities for converting raw binary DAQ data into CSV files for offline analysis. Since the firmware produces different binary layouts for board-side eMMC logging and live DAQ streaming, the host library maintains separate decoding paths for each format. This prevents logged data from being misinterpreted as streamed data and provides a consistent pathway from raw acquisition files to analysis-ready datasets.

\subsubsection{Edge Gateway}
\label{sec:edge_gateway}

For persistent and autonomous deployments, an edge gateway was developed to support local control, data aggregation, and multi-board coordination. The gateway can be implemented on a single-board computer (SBC), such as a Raspberry Pi. In this configuration, the edge device runs a daemon process that maintains board-facing TCP connections and serves as the local acquisition management node. When powered on or reconnected, AEMS boards automatically connect to the edge device, allowing the gateway to manage one or more boards without requiring a continuously active desktop application.

The gateway separates board-level communication from user-level interaction. The daemon maintains persistent board connections, tracks board state, manages data acquisition jobs, stores metadata, and writes captured data to local storage. Users interact with the system through a command-line interface that communicates with the daemon through a local socket, rather than opening independent connections to each board. This design allows long-duration DAQ jobs to run in the background and supports deployments where the acquisition system must operate independently of a desktop computer.

The edge gateway supports live DAQ streaming, board-side eMMC logging, job tracking, metadata generation, and scheduled acquisition. In live streaming mode, data received from the board are written to local storage on the gateway and can be preserved as raw binary files or converted into CSV files for analysis. In eMMC logging mode, data are recorded directly on the board, while the gateway manages command execution, job status, and metadata. This architecture allows the system to function as a compact and autonomous acquisition platform for laboratory studies, industrial monitoring, and long-duration machine data collection.

\begin{figure}[htbp]
    \centering
    \includegraphics[width=0.9\textwidth]{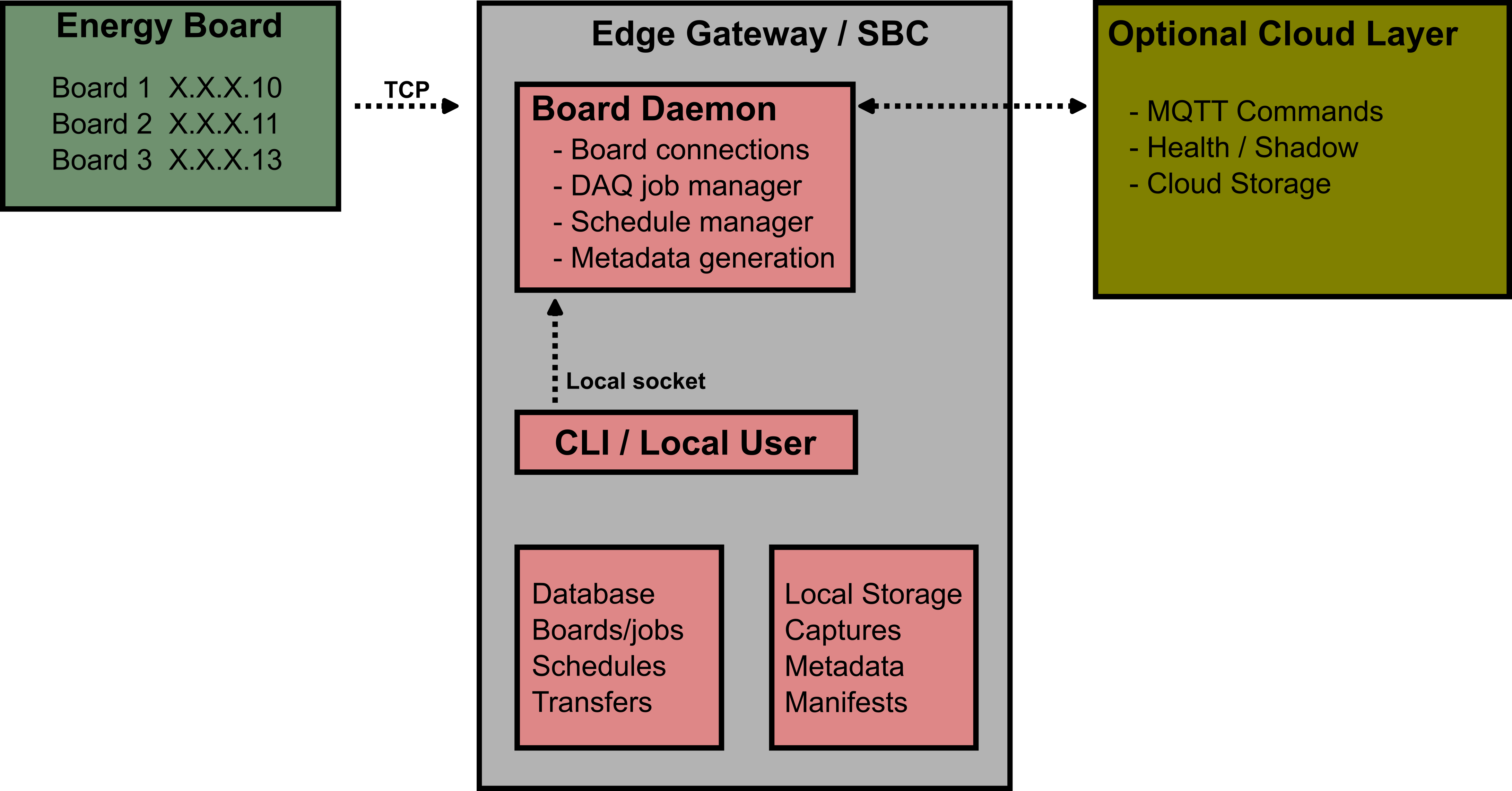}
    \caption{Integrated AEMS edge-gateway architecture. AEMS boards connect to a local edge gateway that manages persistent communication, DAQ jobs, scheduled acquisition, local storage, metadata generation, and transfer manifests. An optional cloud interface provides remote command forwarding, health/shadow reporting, and data transfer while preserving local board control at the gateway.}
    \label{fig:aems_raspberry_pi_gateway}
\end{figure}

The gateway also enables local autonomous operation through scheduled DAQ jobs. Users can define acquisition schedules specifying the target board, acquisition mode, duration, file-name template, sampling rate, channel mask, and block size. These schedules allow the system to perform repeatable data acquisition without manual intervention.

This capability is useful for long-duration machine monitoring, periodic condition monitoring experiments, and unattended industrial deployments. Since schedules and data acquisition jobs are managed locally on the Raspberry Pi, data collection can continue even when a remote host or cloud service is unavailable.

An optional cloud interface was developed to support remote monitoring, command forwarding, health reporting, and data transfer. The cloud interface is implemented as a separate software process that communicates with the local edge device daemon rather than directly controlling the AEMS boards. This local-first design ensures that board communication, scheduled acquisition, and local data capture can continue even if internet connectivity or cloud services are unavailable.

The cloud interface can bridge the local gateway to cloud services such as AWS IoT Core and S3. Through this interface, cloud-issued commands can be forwarded to the local daemon, while board and system status can be published as cloud-compatible health or shadow documents. Captured data, metadata, and transfer manifests can also be uploaded to cloud storage. The manifest-based transfer mechanism provides a structured record of transferred files, including file lists, byte counts, and checksums, enabling traceable downstream data ingestion. The gateway architecture can be seen in Figure \ref{fig:aems_raspberry_pi_gateway}.

Overall, the host interface provides a scalable pathway for deploying the energy monitoring hardware across multiple use cases. The host communication library supports direct scripting and rapid experimentation, the edge gateway supports persistent edge deployment and autonomous acquisition, and the optional cloud interface enables remote monitoring and data transfer for larger distributed monitoring systems.

\begin{table}[!htbp]
\caption{Summary of the host-side, edge, and cloud software components developed for the proposed energy monitoring hardware.}
\label{tab:aems_host_software_summary}
\begin{tabular}{>{\raggedright\arraybackslash}p{0.24\linewidth} >{\raggedright\arraybackslash}p{0.25\linewidth} p{0.41\linewidth}}
\toprule
\textbf{Software component} & \textbf{Deployment target} & \textbf{Primary function} \\
\midrule
Host communication library & Laptop, desktop, or edge device & Provides a high-level software interface for board connection management, command execution, DAQ control, file retrieval, live streaming, response parsing, and binary-to-CSV conversion. \\
Edge gateway daemon & Single-board computer or edge gateway & Maintains persistent board connections, manages DAQ jobs, tracks board and job state, supports scheduled acquisition, generates metadata, and creates transfer manifests. \\
\texttt{aemsctl} command-line interface & Edge gateway & Provides local user access for board status, diagnostics, DAQ streaming, eMMC logging, schedule management, data transfer, and system health monitoring. \\
Local storage and metadata layer & Edge gateway & Stores captured files, job metadata, board history, schedules, transfer jobs, and manifests to support traceable and autonomous data acquisition. \\
Optional cloud agent & Cloud-connected edge gateway & Bridges the local edge gateway to cloud services for remote command forwarding, health/shadow reporting, and transfer of captured data and metadata to cloud storage. \\
\bottomrule
\end{tabular}
\end{table}

\subsubsection{Software Deployment Summary}
\label{sec:soft_deploy}

The overall software design supports multiple levels of deployment. For direct experimentation, users can install the Python interface library on a laptop, desktop, or Raspberry Pi and write custom scripts to configure the board, start data acquisition, stream files, and convert data. For persistent edge deployment, the Raspberry Pi daemon provides continuous board connectivity, local job management, scheduled acquisition, metadata generation, and transfer manifest creation. For remote and autonomous operation, the optional cloud agent provides a pathway for MQTT-based command forwarding, cloud-compatible health reporting, and transfer of captured data to cloud storage.

\begin{figure}[!htbp]
    \centering
    \includegraphics[width=0.9\textwidth]{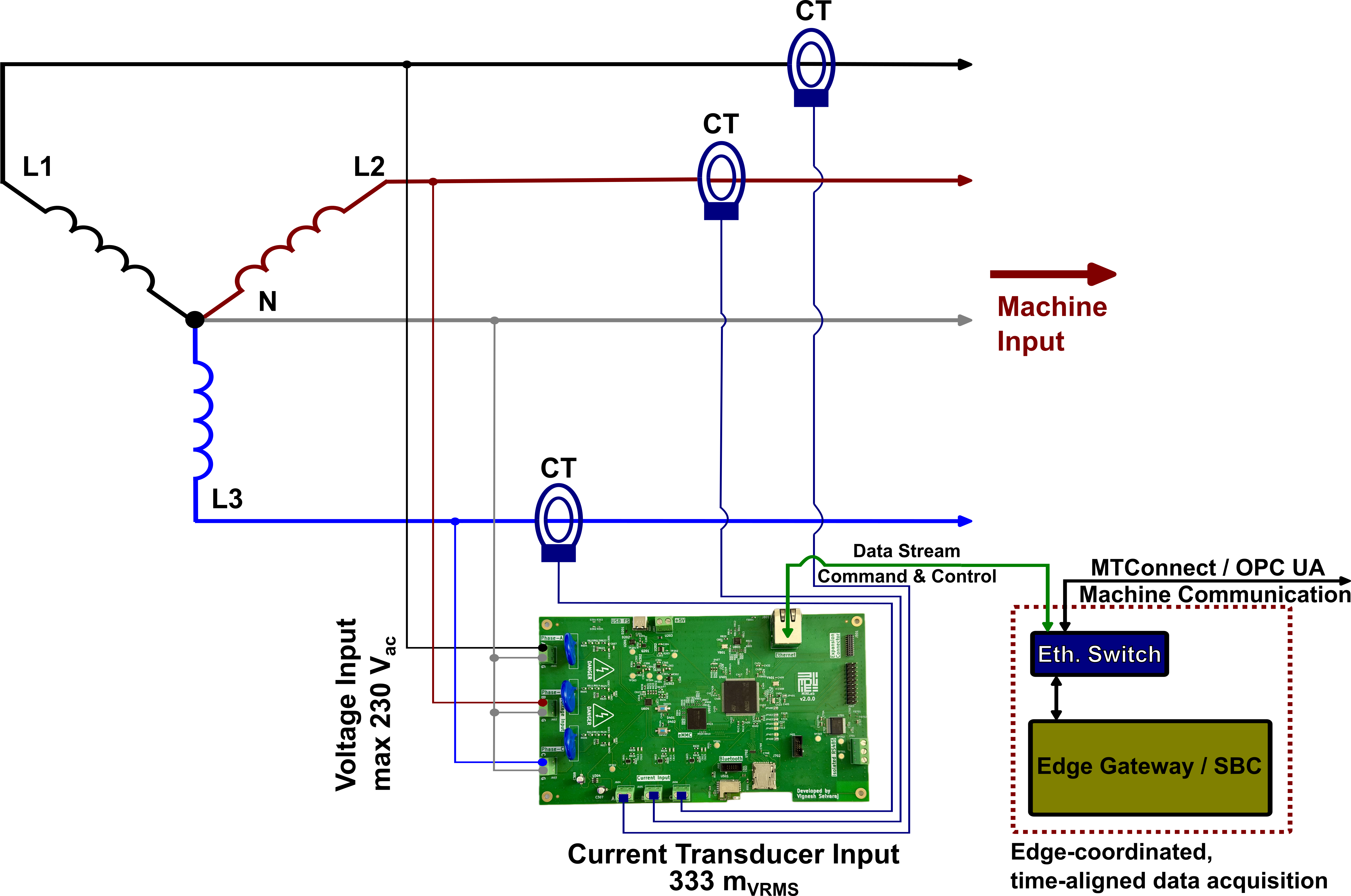}
    \caption{Experimental setup showcasing the connections to the board.}
    \label{fig:exp_board_attach}
\end{figure}

\section{Case study: Deployment and Validation}\label{sec:case_study}
% List of items that can be added here - Subject to change
% 1. Describe the experiments
% - The metal piece along with the G-code path drawn on them
% - A Picture of the setup - Showing the CTs, the connection, and voltage lines etc. Anything beneficial from the experiment standpoint.
% - Details on the experiment conducted.
% - Also indicate where the data is stored.
% - Can we show an actual deployment of the board in industry? Maybe a picture showing that this is how we deployed at an industry for large scale study?
% 2. Insights from the experiment
% - Is this hardware effective? Worth usin
% - The benefits gained are validated
% - Sensitivity study
% - Quantify the increase in energy - I think this falls back into the resolution aspect of the hardware. But can we quantify it or mention it in regards to increase in energy that we can measure. ( Maybe this is a stupid idea)
% Etc.,
% Probably integrated the next section into this as well?

To evaluate the performance of the board on the CNC machine, one board was deployed on a three-axis HAAS CNC machine. The current measurement was performed using the current transformers, while the voltage measurement was performed by directly attaching leads to the three-phase input of the machine. Figure \ref{fig:exp_board_attach} shows the electrical connections to the board.

\begin{figure}[htbp]
    \centering
    \includegraphics[width=0.9\textwidth]{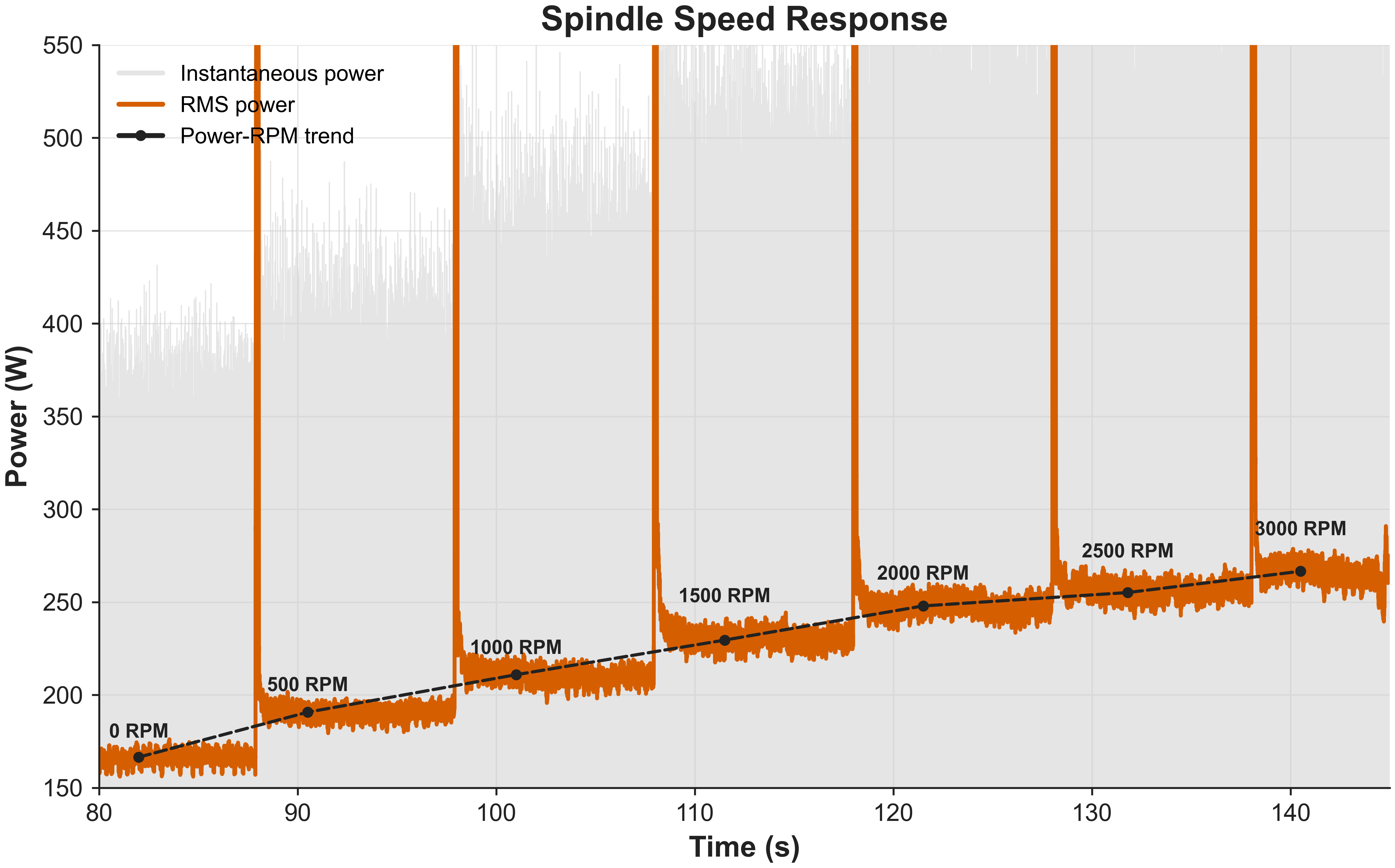}
    \caption{RMS power during air cutting at different spindle speeds.}
    \label{fig:spindle_rpm}
\end{figure}

The first experiment involved varying the spindle speed during air cutting without any other operation on the machine. The purpose of this experiment is to test the board’s ability of detecting the spindle speed change. Spindle speed was increased from 0 to 3000 RPM with an increment of 500 RPM, and there was a 10 second dwell for each spindle speed gradient to capture the steady state power consumption. Figure \ref{fig:spindle_rpm} shows the power consumption with the spindle speed variation. The data in gray color is the raw power consumption data calculated from three-phase current and voltage, to reduce the noise in the raw data, moving Root Mean Square (RMS) was calculated with a window size of 0.1 second, and the data in red color is the power consumption after applying the moving RMS. With the increase in spindle RPM, a non-linear increasing trend of the power consumption can be seen. The rate of increase is higher in the lower RPM zones compared to the higher RPM zones.

\begin{figure}[!htbp]
    \centering
    \includegraphics[width=\textwidth]{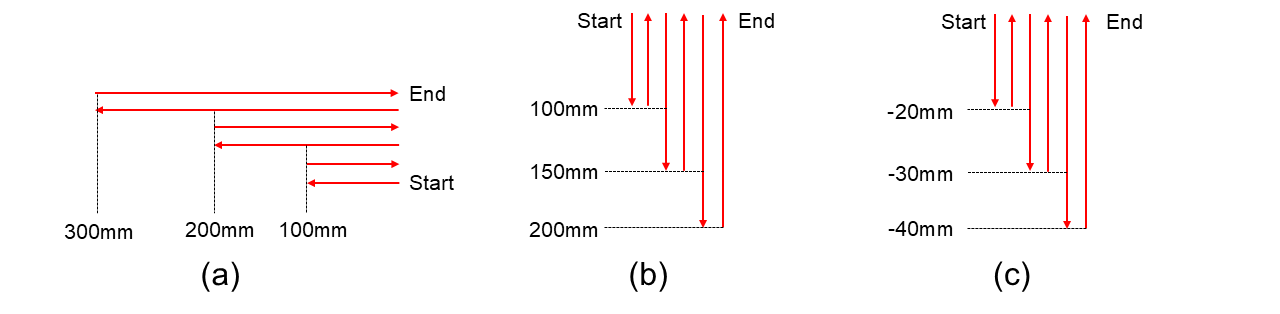}
    \caption{X-, Y-, and Z-axis traverse for the experiment on feed rate variations.}
    \label{fig:xyz_exp_traverse}
\end{figure}

The second experiment examined the effect of feed rate variation under air-cutting conditions with the spindle speed held constant. The spindle was set to 1500 RPM, and the X-, Y-, and Z-axes were commanded at a range of feed rates to evaluate the board's ability to resolve feed-rate-induced changes in power consumption. Figure \ref{fig:xyz_exp_traverse} illustrates the tool paths used in this test, with subfigures (a), (b), and (c) corresponding to motion along the X-, Y-, and Z-axes, respectively. Because motion along the X- and Y-axes is produced by the machine table, whereas motion along the Z-axis is produced by the spindle head, the X- and Y-axis power data were acquired together, while the Z-axis power data were acquired in a separate run.

\begin{figure}[!htbp]
    \centering
    \includegraphics[width=0.9\textwidth]{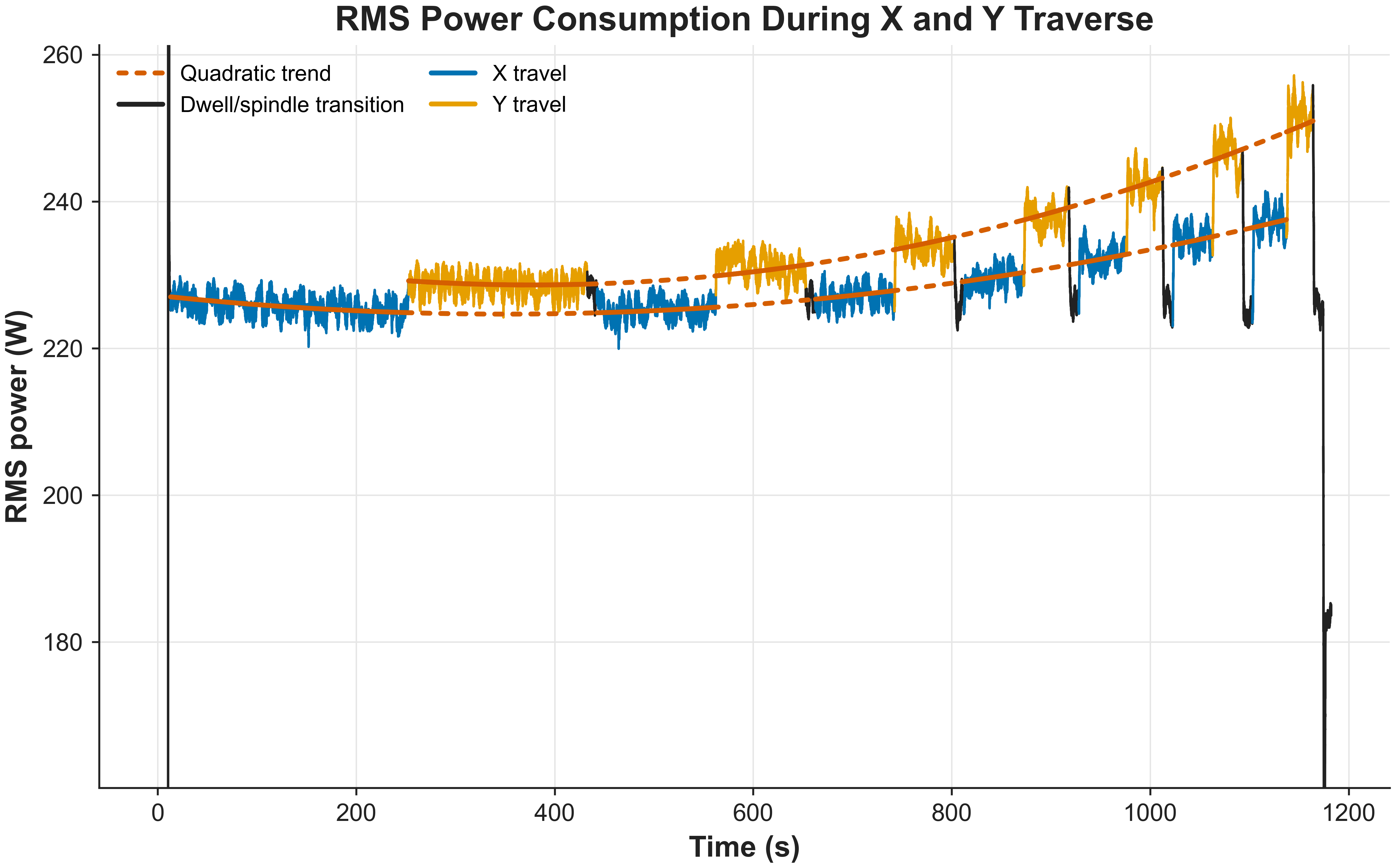}
    \caption{RMS power during air cutting for feed rate variation in X and Y axis, from 300 to 2100 mm/min.}
    \label{fig:feed_aircutting_XY}
\end{figure}

\begin{figure}[!htbp]
    \centering
    \includegraphics[width=0.9\textwidth]{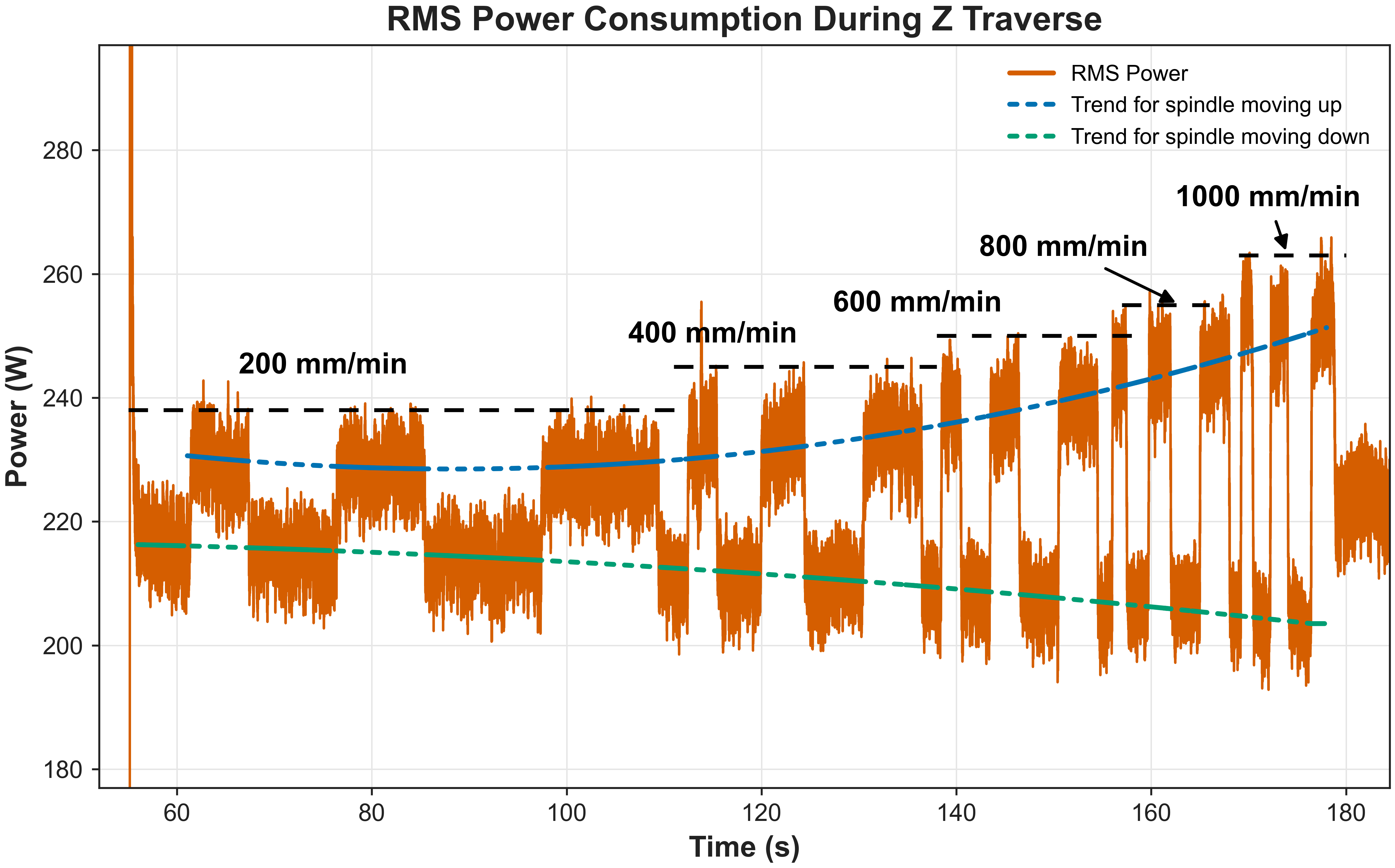}
    \caption{RMS power during air cutting for feed rate variation in Z axis, from 200 to 1000 mm/min.}
    \label{fig:feed_aircutting_Z}
\end{figure}

For the X- and Y-axis trials, the initial feed rate was set to 300 mm/min. The X-axis was first traversed along the tool path shown in Figure \ref{fig:xyz_exp_traverse}a, followed by the Y-axis along the tool path shown in Figure \ref{fig:xyz_exp_traverse}b. The feed rate was then incremented in 300 mm/min steps up to 2100 mm/min, with the same tool paths executed at each step. A 10 s dwell was programmed between successive feed-rate increments to ensure clear delineation of steady-state regions in the recorded signal. Figure \ref{fig:feed_aircutting_XY} presents the RMS power consumption corresponding to the feed-rate sweep along the X- and Y-axes. Power consumed during X-axis motion is shown in blue, power consumed during Y-axis motion is shown in orange, and dwell regions are shown in black. The dwell and axis-motion segments were labeled by cross-referencing the recorded power signal with synchronized video of the machine's control panel (G-code execution, X/Y position readout, and dwell countdown). The results indicate that power consumption along both the X- and Y-axes increases monotonically with feed rate, exhibiting a nonlinear trend that approaches an exponential growth profile over the tested range. In addition, Y-axis traverses consistently drew more power than X-axis traverses at equivalent feed rates, which is consistent with the larger inertial load carried by the Y-axis (saddle plus table) compared with the X-axis (table only) in a typical three-axis vertical machining center configuration. For the case of the Z-axis, the spindle speed was set to 1500 RPM, and an initial feed rate of 200 mm/min. The feed rate was then increased to 1000 mm/min in increments of 200 mm/min, and in each level the Z-axis moved in the path shown in Fig \ref{fig:xyz_exp_traverse}c. Figure \ref{fig:feed_aircutting_Z} shows the RMS power for feed rate variation in Z-axis. It can be seen that the board can successfully capture the interlaced pattern of the power consumption for Z-axis traverse, it can be clearly observed that spindle moving down consumed less power due to gravity. Additionally, board can capture the power consumption for spindle moving up with different feed rates, and an increasing trend can be identified. 

\begin{figure}[!htbp]
    \centering

    \begin{subfigure}[b]{0.9\textwidth}
        \centering
        \includegraphics[width=\textwidth]{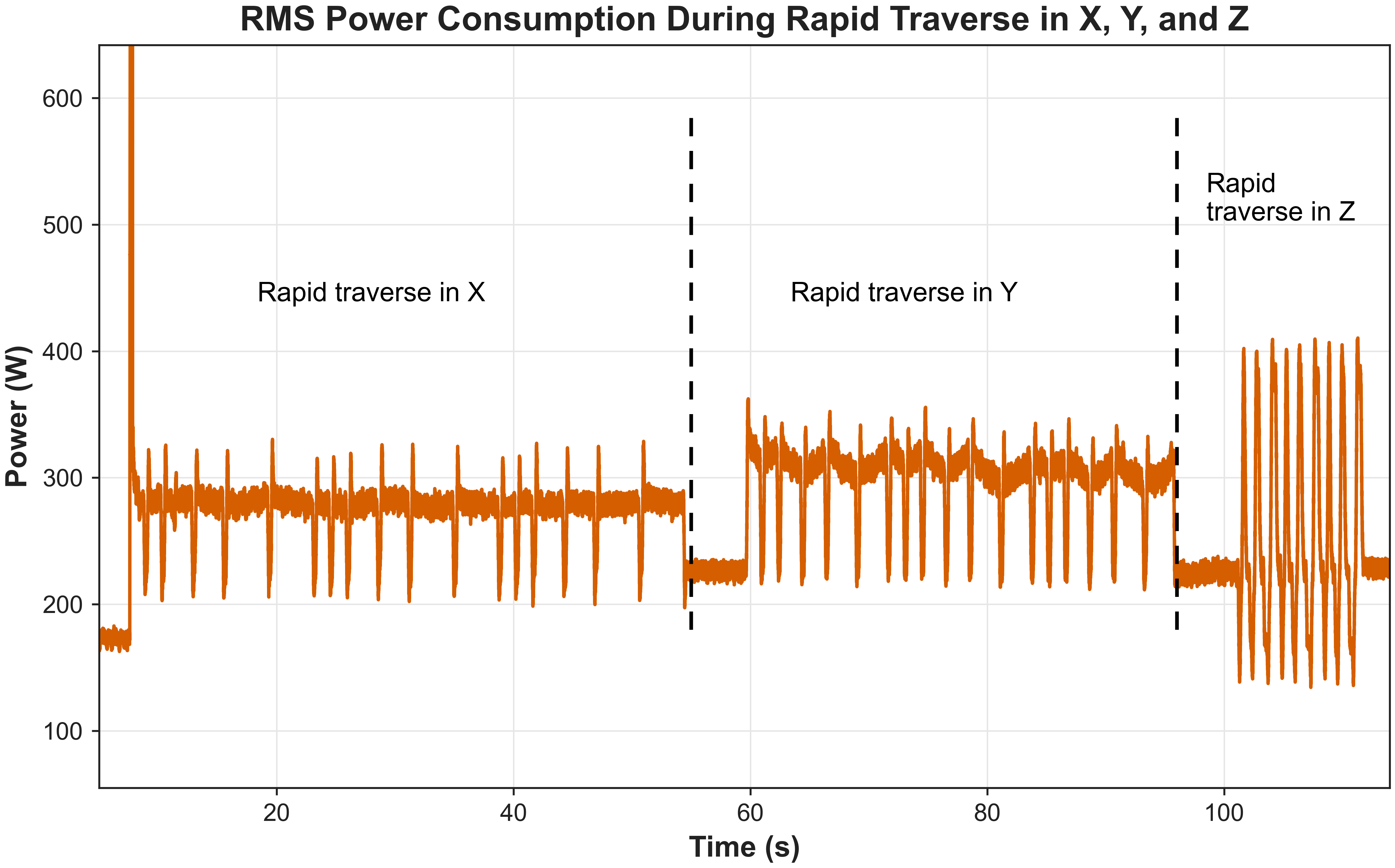}
        \caption{RMS power during rapid traverse on X-, Y-, and Z-axis.}
        \label{fig:rapid_traverse_xyz}
    \end{subfigure}

    \vspace{0.5em}

    \begin{subfigure}[b]{0.9\textwidth}
        \centering
        \includegraphics[width=\textwidth]{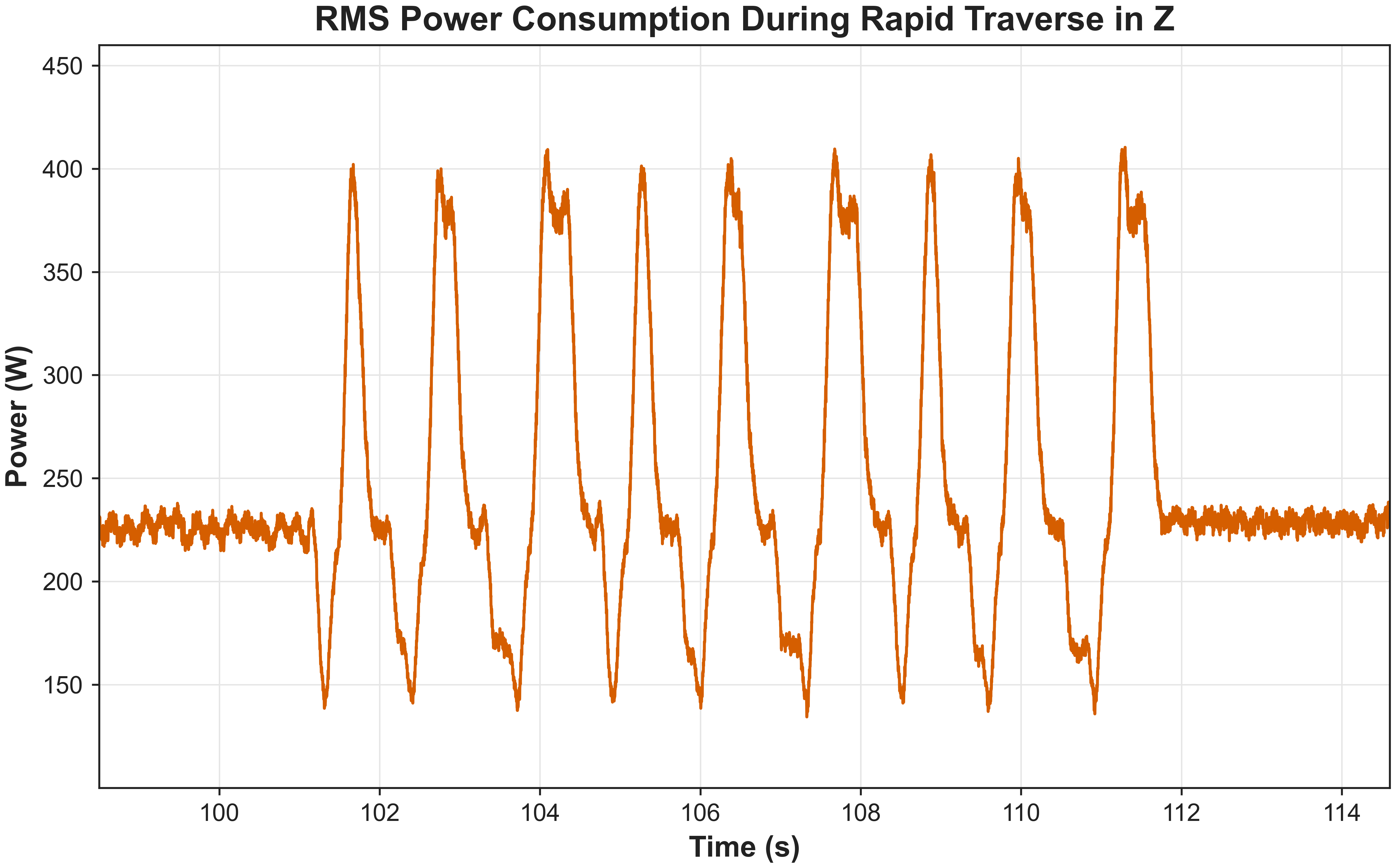}
        \caption{RMS power during rapid traverse on Z-axis.}
        \label{fig:rapid_traverse_z}
    \end{subfigure}

    \caption{RMS power during rapid traverse motion.}
    \label{fig:rapid_traverse}
\end{figure}

The third experiment examined the power consumption associated with rapid traverse moves under air-cutting conditions. The spindle speed was held at 1500 RPM, and the X-, Y-, and Z-axes were commanded along the same tool paths used in Figure \ref{fig:xyz_exp_traverse}. Each axis was cycled through its respective path three times, producing a total of 18 linear rapid-traverse moves per axis. Figure \ref{fig:rapid_traverse_xyz} shows the aggregate RMS power signal captured during the rapid-traverse sequence, in which all 18 moves per axis are clearly resolved. Figures \ref{fig:rapid_traverse_z} present the corresponding RMS power profile for the Z-axis, providing a finer view of the per-axis behavior. In Figure \ref{fig:rapid_traverse_xyz}, the large initial peak corresponds to the spindle start-up transient prior to the traverse sequence. In Figure \ref{fig:rapid_traverse_z}, the downward excursions of the power consumption profile correspond to Z-axis descent (head moving toward the table) and the upward excursions correspond to Z-axis retraction.

\begin{figure}[!htbp]
    \centering
    \includegraphics[width=0.95\textwidth]{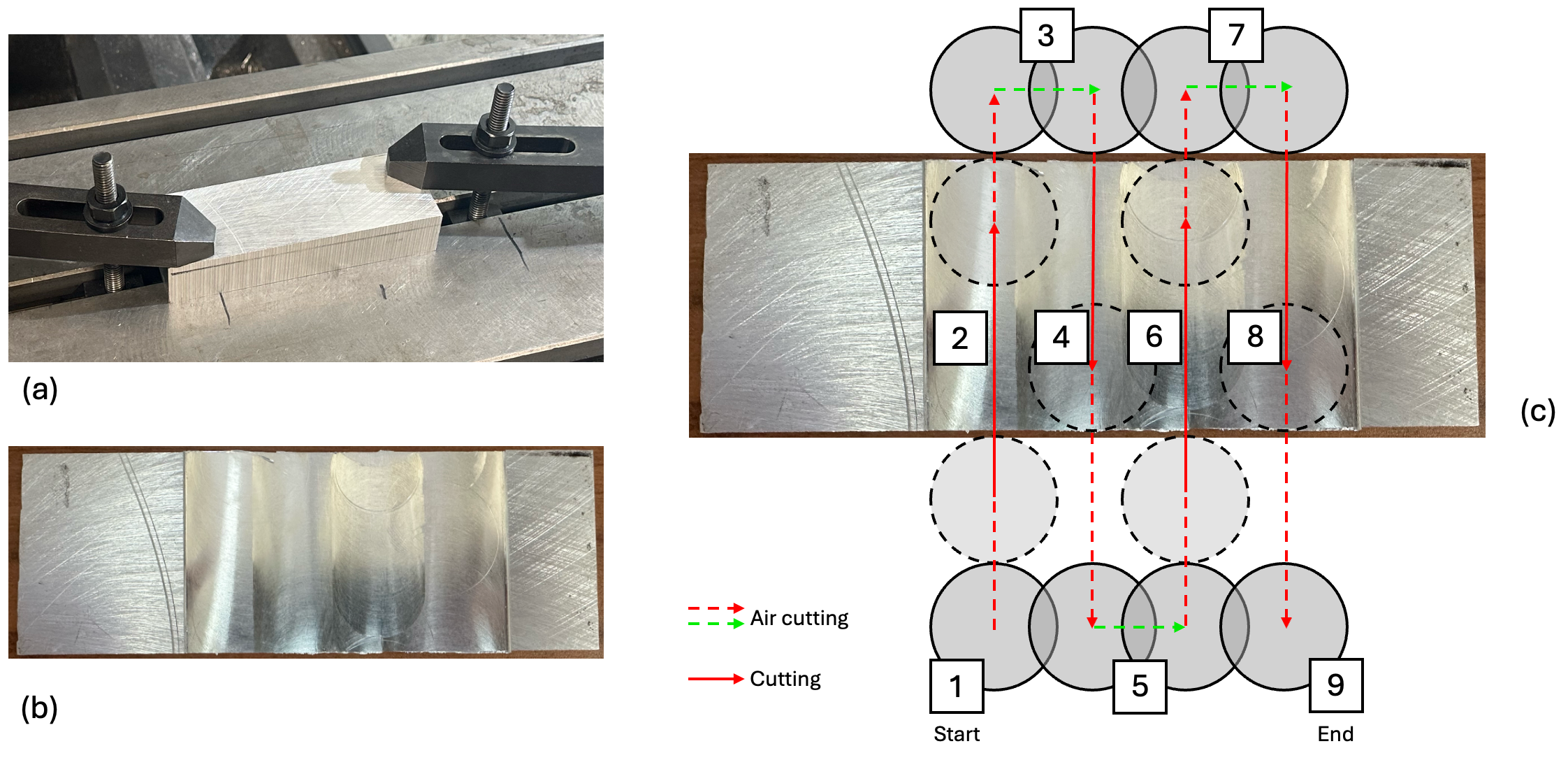}
    \caption{Workpiece setup for energy monitoring during machining operation.}
    \label{fig:wp_setup}
\end{figure}

\begin{figure}[!htbp]
    \centering
    \includegraphics[width=0.9\textwidth]{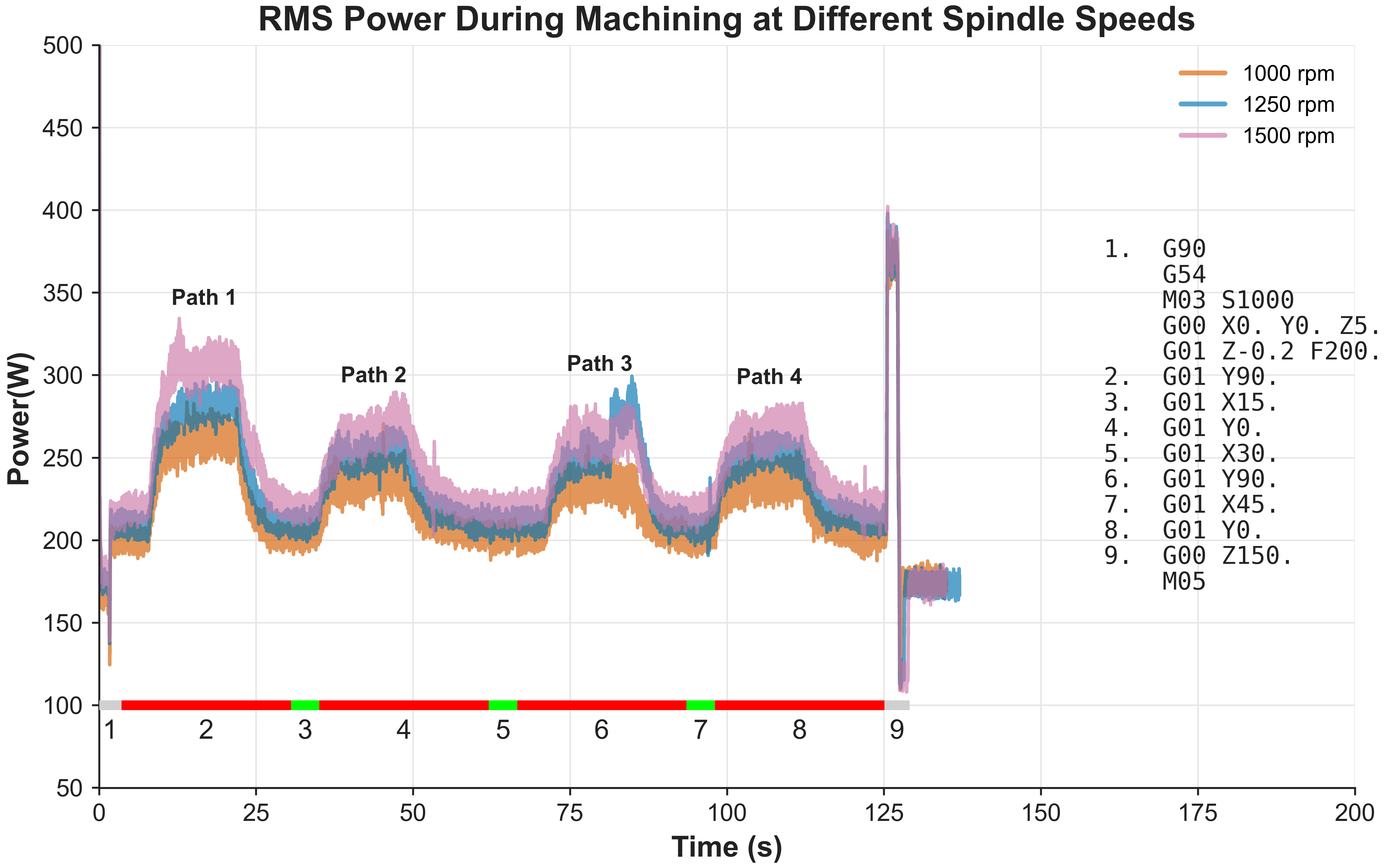}
    \caption{RMS power consumption during face-milling investigating the effect of spindle rpm. The power consumption patterns were correlated with the gcode governing the machine operation.}
    \label{fig:cutrpm}
\end{figure}

\begin{figure}[!htbp]
    \centering
    \includegraphics[width=0.9\textwidth]{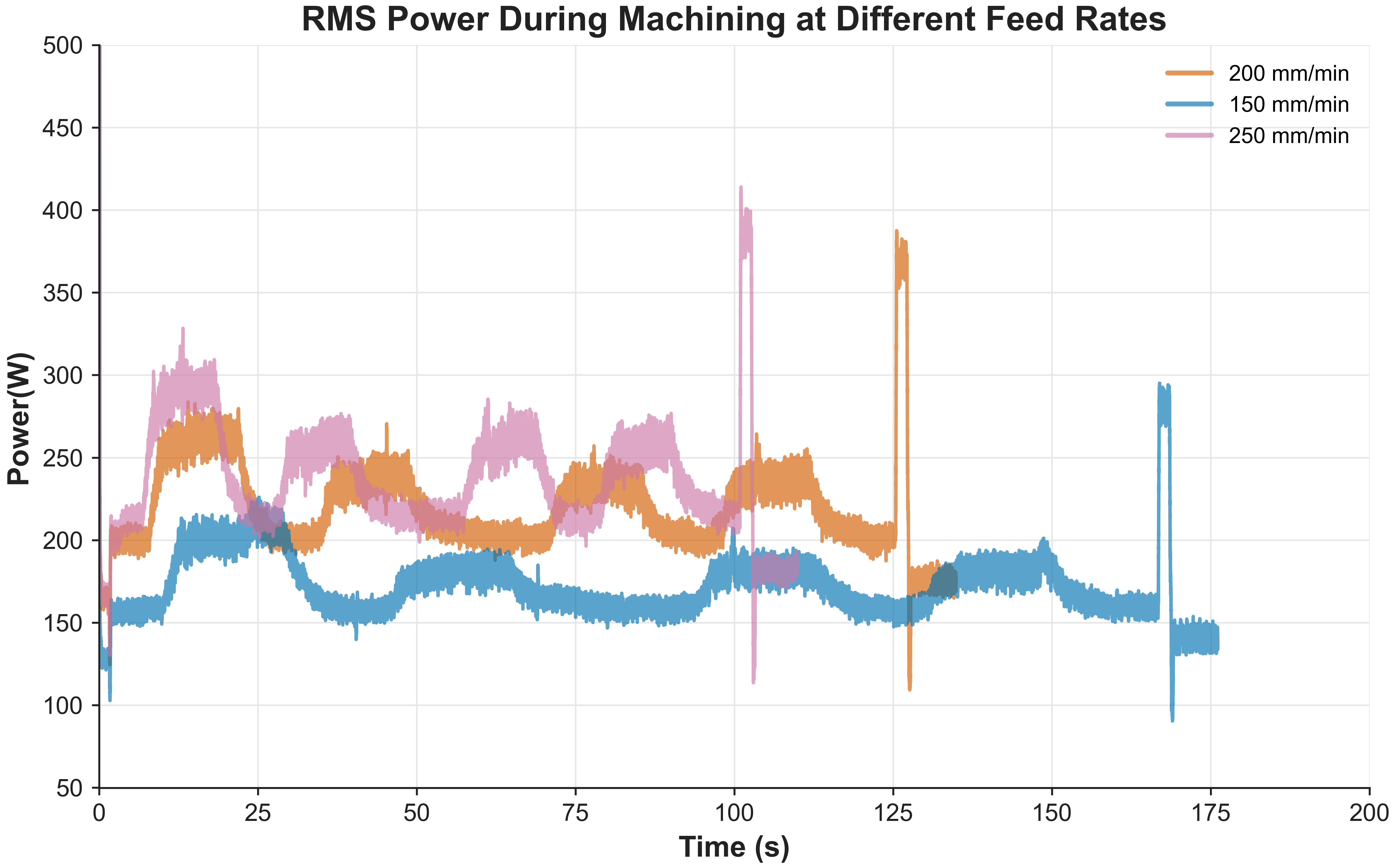}
    \caption{RMS power consumption during face-milling investigating the effect of feed rate.}
    \label{fig:cutfeed}
\end{figure}

The final experiment evaluated the board's ability to capture cutting-energy signatures across different cutting conditions. A face-milling operation was performed on a rectangular aluminum workpiece. Figures \ref{fig:wp_setup}a and \ref{fig:wp_setup}b show the workpiece before and after machining, respectively, and Figure \ref{fig:wp_setup}c shows the tool path programmed on the workpiece surface. Cutting fluid was disabled throughout the experiment to eliminate coolant-pump power draw as a confounding variable in the recorded signal. The first part of the experiment investigated the effect of spindle speed. The spindle speed was stepped from 1000 RPM to 1250 RPM and then to 1500 RPM between successive cutting passes, while all other cutting parameters were held constant. Figure \ref{fig:cutrpm} shows the RMS power consumption recorded during the spindle-speed sweep. For the tool path shown in Figure \ref{fig:wp_setup}c, the cutter engages the workpiece four times per pass, and all four tool-workpiece engagement events are clearly resolved in the recorded signal. As expected, the machining power consumption increases with spindle speed. The second part of the experiment investigated the effect of feed rate. The feed rate was stepped from 100 to 150 and then to 200 mm/min between successive cutting passes, with the spindle speed and all other cutting parameters held constant. Figure \ref{fig:cutfeed} shows the RMS power consumption across the three feed rates. Power consumption increases monotonically with feed rate, as expected from the corresponding increase in material removal rate, while the overall shape of the power profile is preserved across passes because the programmed tool path is identical. These results demonstrate that the board is able to resolve feed-rate variations as small as 50 mm/min during machining.

% 1. Give the tool-workpiece combination - The range of RPM is selected based on this combination - We need to say that the hardware is able to resolve that.
% 2. For Path 3 in blue, there is a spike what can be the reason for this spike?

\section{Discussion}\label{sec:discussion}
% Explaining the results from the experiment and discussions
% Discussion
% - Impact of environmental factors on the sensors (CIRP Paper CP)
% - FPGA for edge quick data processing
% - Integrating sensor with parts? Is this possible in our case
% Convince the audience why we need this, everything is so scattered, we need a way to bring the community together to enable this process of sensor.daq systekm development

The case study in Section 4 establishes that AEMS resolves the principal energy states of a three-axis machining center such as spindle no-load behavior, feed-drive motion, rapid traverse, and material removal, with sufficient fidelity to distinguish them from one another and to track parametric changes as small as a 50 mm/min feed-rate step. These measurements, however, are only the immediate evidence of the system's performance. The broader contribution of this work lies not in the specific power signatures recorded on a single machine, but in demonstrating that such signatures can be captured on an open, low-cost, and reproducible platform that any laboratory or small manufacturer can rebuild and extend. In addition, the system's ability to perform edge-coordinated, time-aligned acquisition enables large-scale studies in real industrial settings, with optional remote management through cloud services. This section situates the system within the design constraints that motivated it, the present state of open-source hardware in manufacturing, and the research gap that a reproducible energy-acquisition platform is positioned to close.

\subsection{Rationale for a custom hardware design}

The decision to develop a dedicated board, rather than adopt an existing instrument, followed from the observation that no single class of available hardware satisfies the combined requirements of research-grade fidelity, industrial robustness, openness, and low cost set out in Section 3.1. Three established options were considered, and each fails at least one of these requirements.

Commercial power and energy meters offer high accuracy and certified measurement, but they are designed predominantly for aggregated power monitoring and billing rather than for transient analysis. As noted by \cite{o2013industrial, kara2011electricity}, such instruments are typically proprietary, operate at sampling rates suited to facility- or process-level accounting, and restrict access to the raw waveform data that machine-level diagnostics and data-driven modeling require. Their closed architectures and per-unit cost further preclude the dense, fleet-scale instrumentation needed to correlate energy events across machines.

Laboratory-grade data acquisition systems deliver the necessary sampling fidelity, but at a cost and in a form factor that scale poorly. Such systems are oriented toward bench experimentation rather than permanent installation, and their rigidity and price make plant-wide deployment economically impractical — the same trade-off between monitoring capability and affordability that \cite{o2012implementation} identify as the principal barrier to wider adoption.

At the opposite end, general-purpose open hardware platforms such as Arduino and Raspberry Pi have driven much of the experimentation in IoT and IIoT applications, \cite{krushinitskiy2013review}, and several industrial variants now exist. Their use in industrial measurement nonetheless remains contested. As \cite{martikkala2021trends} observe, the barriers to adopting open solutions for the IIoT are not primarily computational but concern interoperability, standardization, reliability, and regulatory compliance. General-purpose boards lack the isolated analog front-end, simultaneous-sampling conversion, surge protection, and industrial communication interfaces required for trustworthy energy measurement on a factory floor.

AEMS was therefore designed to occupy the intersection that none of these options reaches. Each design choice described in Section 3.2 answers a specific shortcoming of the alternatives: the 24-bit, simultaneous-sampling ADC and the isolated analog and digital rails provide the fidelity withheld by commercial meters; the surge protection and industrial form factor provide the robustness absent from general-purpose boards; the RS-485/Modbus and Ethernet interfaces address the interoperability and standardization concerns raised by \cite{martikkala2021trends}; and on-board eMMC logging preserves high-rate acquisition where network bandwidth is limited. Openness and low cost remove the access barrier that the laboratory and commercial options impose, and finally, all the design choices were made with large scale deployment and studies for AI applications.

\subsection{Open-source hardware and the gap in industrial energy monitoring}

Although open-source hardware (OSH) has a well-documented record of reducing cost and accelerating innovation, its penetration into manufacturing remains limited. \cite{stange2023open} report OSH adoption in the manufacturing domain to be as low as 4.6\%, with activity concentrated in the hobbyist community, even as \cite{pearce2014cut, pearce2020economic} documents savings of up to 87\% relative to proprietary scientific instrumentation. This disparity points to a gap that is structural rather than economic: the open platforms that are accessible are not industrially credible, while the instruments that are industrially credible are neither open nor affordable at scale.

The consequences of this gap are most acute in energy monitoring. Manufacturing accounts for roughly 90\% of the industrial sector's energy consumption and 84\% of its energy-related CO2 emissions \citep{duflou2012towards}, yet the high-resolution, multi-level visibility that meaningful efficiency improvements require \citep{rahimifard2010minimising} depends on instrumentation that most facilities, particularly small and medium-sized enterprises (SMEs), cannot justify deploying densely. The result is that the machines most in need of monitoring are frequently the least instrumented.

Beyond cost, openness carries a distinct value for the research community: it shortens the knowledge-transfer cycle from universities to industry. Collaborative OSH projects allow research prototypes to converge on industrial requirements more rapidly and offer SMEs, which rarely maintain in-house research capacity, a route to innovation through shared designs \citep{sedita2016role}. Viewed in this light, an open energy-acquisition platform is not merely a less expensive meter; it is shared research infrastructure.

\subsection{Filling the gap through a reproducible, open platform}

The platform presented in this work is positioned to close this gap along four dimensions.

First, reproducibility. Because the complete stack — hardware schematics, embedded firmware, and the host, edge, and cloud software — is released openly, the platform can be rebuilt, audited, and extended rather than only cited. Reproducibility is the property that distinguishes shared infrastructure from a one-off prototype, and it is a precondition for the cumulative, comparable measurements that data-driven manufacturing research depends upon.

Second, democratization of access. The combination of low cost and openness lowers the barrier to entry so that SMEs and research laboratories, and not only large original-equipment manufacturers, can instrument their machines at the resolution required for diagnostics. In doing so, the platform directly addresses the technological divide between large corporations and smaller enterprises identified in Section 1.

Third, scalability to the plant level. The synchronous multi-board acquisition described in Section 3 extends monitoring from a single machine to coordinated, plant-wide measurement. This capability is what enables the cross-machine correlation, interaction analysis, and bottleneck identification that motivate energy monitoring beyond the level of an individual asset.

Fourth, the enabling of downstream analytics. Raw, high-rate, synchronized data is the substrate that AI/ML condition monitoring, digital twins, and non-intrusive load monitoring require but that conventional metering withholds. The case study provides direct evidence that the acquired data carries this information: the system distinguishes which axis is in motion from power alone and resolves cutting signatures down to small feed-rate increments. Moreover, the observed trends are interpretable within established theory. The monotonic increase of cutting power with feed rate, and hence with material removal rate, is consistent with the functional relationship between energy consumption and material removal rate established by \cite{gutowski2006electrical}. The platform's output is therefore not merely raw signal, but data that maps onto existing energy models and can support the refinement of existing energy models.

\subsection{System Limitations}

Several limitations bound the present results and define the immediate scope for further work. The validation reported here is qualitative: the system is shown to resolve known machine events and parametric changes, but it has not yet been benchmarked against a calibrated reference power analyzer. The case study is also confined to a single machine, tool, and workpiece material; although synchronous multi-board acquisition is implemented and described, it has not yet been demonstrated experimentally across multiple machines. Finally, the measurement envelope, three-phase input to 230 VAC, 32 kS/s per channel, and 333 mV current transformers, bounds the range of equipment the board can presently monitor, excluding, for example, higher-power spindles and the harmonic content introduced by variable-frequency drives, and the linearity of the current transformers at low load warrants further investigation. These limitations are addressed in the directions outlined in Section 6.

\section{Conclusion and Future Work}\label{sec:conclusion}

This work was motivated by a structural gap in industrial energy monitoring: the instruments that are industrially credible are proprietary, costly, and closed to the raw high-frequency data that modern analytics require, while the open and affordable platforms that could democratize measurement lack the front-end fidelity, isolation, and robustness needed for trustworthy deployment on a factory floor. Neither class of hardware sits at the intersection of research-grade fidelity, industrial robustness, openness, and low cost, and the absence of an option that does, has kept high-resolution energy data out of reach for the small and medium-sized enterprises and research laboratories that stand to benefit from it most.

To address this gap, we presented Autonomous Energy Monitoring System (AEMS), an open, low-cost, and reproducible platform for high-resolution energy data acquisition in manufacturing. The system spans the full acquisition stack: a sensing and measurement front-end built around a 24-bit, simultaneous-sampling analog-to-digital converter with isolated power domains and surge protection; a dual-core processing core that separates deterministic acquisition and storage from host communication and control; and a user-friendly software stack comprising a host communication library, an edge-gateway daemon, and an optional cloud interface that together support deployment from direct laboratory experimentation to autonomous, plant-scale monitoring. Industrial communication interfaces and synchronous multi-board acquisition were incorporated to meet the interoperability and scalability requirements that distinguish a research instrument from a deployable one.

The system was validated in a case study on a three-axis Haas vertical machining center, in which it resolved the principal energy states of the machine, spindle no-load behavior, feed-drive motion, rapid traverse, and material removal, and tracked parametric changes as fine as a 50 mm/min feed-rate step. The recorded signatures were not only distinguishable from one another but also interpretable within established machine-tool energy theory, with cutting power increasing monotonically with feed rate in a manner consistent with the energy–material-removal-rate relationship described in the literature. Together, these results demonstrate that research-grade energy signatures can be captured on hardware that any laboratory or manufacturer can rebuild and extend. The complete hardware design and firmware are released openly at \url{https://github.com/vigneshuw/aems} and archived on Zenodo \citep{aems_software}, so that the platform can serve as shared research infrastructure rather than a single-use prototype.

Several directions remain for future work, following directly from the limitations identified in Section 5.4. The most immediate is quantitative validation: benchmarking the system against a calibrated reference power analyzer to characterize gain and phase error, total harmonic distortion, and inter-channel timing skew, and thereby establish formal accuracy and uncertainty bounds. A second direction is the experimental demonstration of synchronous multi-board acquisition across multiple machines, extending the single-machine results reported here to the plant-level, time-aligned monitoring the architecture was designed to support. A third is broadening the measurement envelope to accommodate higher-power spindles and the harmonic content introduced by variable-frequency drives, together with a closer examination of current-transformer linearity at low load. Finally, the high-rate, synchronized data the platform produces opens a path toward on-edge machine-learning inference, for tool-wear classification, anomaly detection, and automatic segmentation of energy states, and toward integration with machine-context streams such as MTConnect and OPC UA, so that energy signatures can be labeled with the process events that produce them. Pursued together, these directions would advance AEMS from a demonstrated acquisition platform toward a complete, open foundation for intelligent energy and process monitoring in Industry 4.0.

\section*{Code and Data Availability}
The complete AEMS design is openly available. The hardware (KiCad schematics, PCB layout, bill of materials, and manufacturing files), the STM32H745 dual-core firmware, and the host-side Python control software are hosted at \url{https://github.com/vigneshuw/aems} and archived on Zenodo \citep{aems_software}. The hardware is licensed under CERN-OHL-P-2.0, the firmware and software under MIT, and the documentation under CC-BY-4.0.

\section*{Acknowledgement}

The authors of this work acknowledge the MAUM (Manufacturing Advancement through Unprecedented Morphing) Consortium at MINLab (Manufacturing Innovation Network Laboratory) for their support in developing the board and enabling test cases in industries.

\bibliographystyle{unsrtnat}
\bibliography{references}% common bib file

\end{document}